\documentclass[9pt,sigconf,letterpaper,natbib=false]{acmart}

\AtBeginDocument{%
  \providecommand\BibTeX{{%
    \normalfont B\kern-0.5em{\scshape i\kern-0.25em b}\kern-0.8em\TeX}}}

\usepackage{booktabs}
\usepackage{multirow}

\usepackage{svg}

\usepackage{tcolorbox}
\usepackage{enumitem}

\usepackage{soul}

\newcommand{\roundframe}[1]{{\setlength\fboxrule{0pt}\fbox{\tcbox[colframe=black,colback=white,shrink tight,boxrule=0.5pt,extrude by=2.5pt]{\small #1}}}}

\setcopyright{rightsretained}
\acmYear{2024}\copyrightyear{2024}
\acmConference[MOBISYS '24]{The 22nd Annual International Conference on Mobile Systems, Applications and Services}{June 3--7, 2024}{Minato-ku, Tokyo, Japan}
\acmBooktitle{The 22nd Annual International Conference on Mobile Systems, Applications and Services (MOBISYS '24), June 3--7, 2024, Minato-ku, Tokyo, Japan}
\acmDOI{10.1145/3643832.3661883}
\acmISBN{979-8-4007-0581-6/24/06}

\acmBadge[https://www.acm.org/publications/policies/artifact-review-and-badging-current]{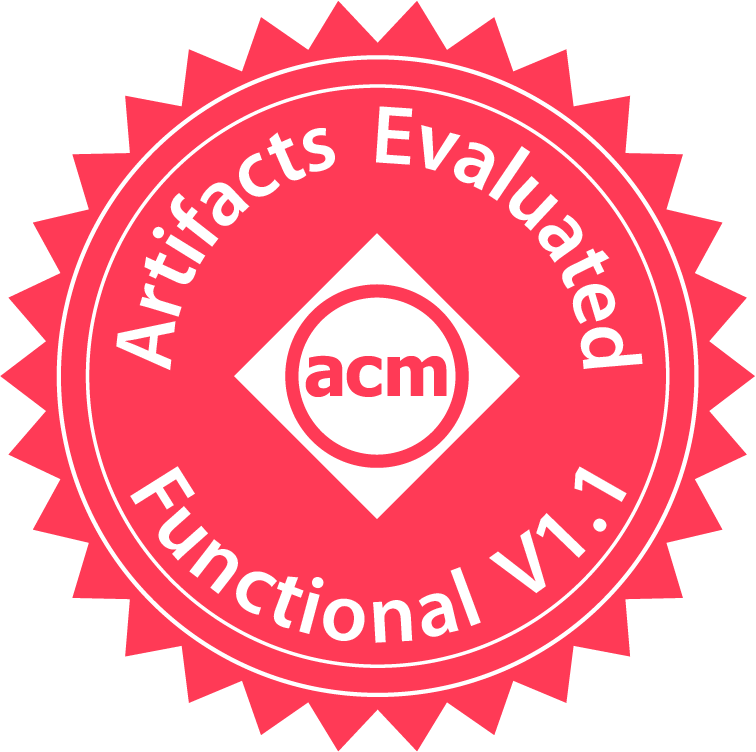}
\acmBadge[https://www.acm.org/publications/policies/artifact-review-and-badging-current]{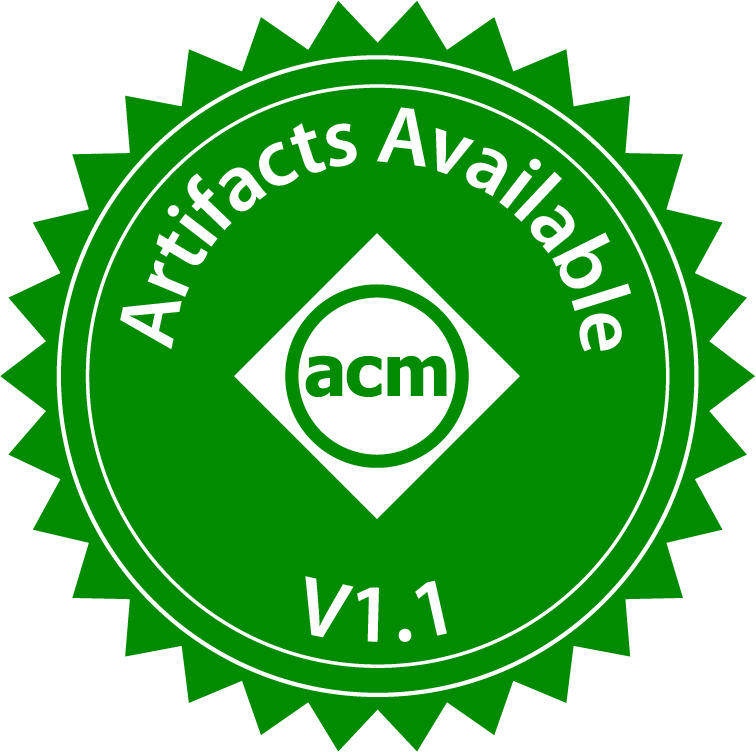}

\RequirePackage[
  datamodel=acmdatamodel,
  style=acmnumeric,
  ]{biblatex}

\begin{document}

\title{
Why E.T. Can't Phone Home:\\
A Global View on IP-based Geoblocking at VoWiFi
}

\author{Gabriel K. Gegenhuber}
\email{gabriel.gegenhuber@univie.ac.at}
\affiliation{%
  \institution{University of Vienna}
  \department{Faculty of Computer Science}
  \department{Doctoral School Computer Science}
  \city{Vienna}
  \country{Austria}
}

\author{Philipp É. Frenzel}
\email{pfrenzel@sba-research.org}
\affiliation{%
  \institution{SBA Research}
  \city{Vienna}
  \country{Austria}
}
\author{Edgar Weippl}
\email{edgar.weippl@univie.ac.at}
\affiliation{%
  \institution{University of Vienna}
  \department{Faculty of Computer Science}
  \city{Vienna}
  \country{Austria}
}
\renewcommand{\shortauthors}{Gegenhuber et al.}

\begin{abstract}

In current cellular network generations (4G, 5G) the IMS (IP Multimedia Subsystem) plays an integral role in terminating voice calls and short messages.
Many operators use \mbox{VoWiFi} (Voice over Wi-Fi, also Wi-Fi calling) as an alternative network access technology to complement their cellular coverage in areas where no radio signal is available (e.g., rural territories or shielded buildings).
In a mobile world where customers regularly traverse national borders, this can be used to avoid expensive international roaming fees while journeying overseas, since VoWiFi calls are usually invoiced at domestic rates.
To not lose this revenue stream, some operators block access to the IMS for customers staying abroad.

This work evaluates the current deployment status of VoWiFi among worldwide operators and analyzes existing geoblocking measures on the IP layer by measuring connectivity from over 200 countries.
We show that a substantial share (IPv4: 14.6\%, IPv6: 65.2\%) of operators implement geoblocking at the DNS- or VoWiFi protocol level, and highlight severe drawbacks in terms of emergency calling service availability.

\end{abstract}


\settopmatter{printfolios=true}

\maketitle
\title{
Why E.T. Can't Phone Home:
A Global View on IP-based Geoblocking at VoWiFi
}

\newcommand{\parvspace}{\vspace{0.20cm}}

\section{Introduction}

Mobile network services are a crucial lifeline in today's society, given that in 2023 over 5.4 billion people relied on cellular networks for connectivity and communication~\cite{GSMA2023MobileEconomy}.
With 4G currently being the most used wireless standard
and 5G rapidly gaining penetration,
numerous operators are actively decommissioning older legacy networks (2G and 3G), marking the completion of the shift from circuit-switched to a comprehensive packet-switched network paradigm.

In the packet-switched domain, operators use VoIP (Voice over IP) based technology to terminate voice calls and messages.
Additionally to the VoLTE (Voice over LTE) standard, VoWiFi (Voice over Wi-Fi, also known as Wi-Fi calling) was introduced.
While VoLTE uses the traditional radio infrastructure that is provided by the operator as its access medium, VoWiFi is a complementary solution that allows the use of third-party wireless networks as an alternative uplink to the operator.
Consequently, customers can leverage existing Wi-Fi access points (APs) and continue utilizing their mobile phones for voice calls in areas with poor or no cellular reception.

\begin{figure}[b!]
  \centering
  \fbox{\includegraphics[width=.9\linewidth]{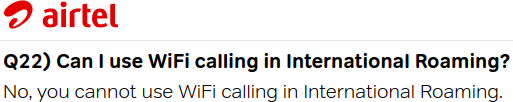}}
  \caption{An Indian operator states in the FAQs~\cite{Airtel2023Blocking} that \mbox{VoWiFi} cannot be used during international roaming.}
  \label{fig:airtel-vowifi-roaming}
\end{figure}

To support this functionality, operators need to expose parts of their infrastructure to the public Internet.
This opens new possibilities for active measurement studies since it allows the investigation of exposed parts of a mobile network without requiring any radio equipment. %
Moreover, it allows measuring a huge number of international operators, without the need for sophisticated measurement hardware at the target locations.

Presumably, the general idea behind VoWiFi is to expand the cellular coverage to allow uninterrupted service e.g., in rural areas with weak reception.
Thereby, a voice call can be handed over from VoLTE to VoWiFi, and vice versa, on the fly.
However, VoWiFi can also be used completely independent from VoLTE, i.e., it requires no radio signal at all and also works e.g., when the mobile phone is in airplane mode but has Wi-Fi connectivity.
In a mobile world that facilitates seamless transitions across national borders, it thereby can also be used overseas, possibly allowing customers to escape from expensive roaming fees.
In practice, some operators are actively denying their customers access to VoWiFi from foreign countries, as the screenshot in Figure~\ref{fig:airtel-vowifi-roaming} shows.

This paper aims to offer a comprehensive overview of the global deployment of VoWiFi and analyzes geoblocking measures of worldwide operators.
More specifically, we use commercial VPN- and cloud services to simulate customers connecting from a diverse set of distinct foreign locations and to thereby determine geoblocking measures.

In summary, the main contributions of this paper are as follows:
\begin{itemize}
    \item We propose a methodological approach to discover existing VoWiFi deployments and to probe them for IP-based geoblocking measures.
    \item We map the current VoWiFi support at worldwide operators, analyze the used infrastructure and give an overview of the latest global market penetration.
    \item We probe worldwide operators for IP-based geoblocking both at the DNS- and VoWiFi-protocol levels and provide an overview of current practices.
\end{itemize}

The remainder of the paper is organized as follows.
Section~\ref{sec:background} introduces the topic by providing background knowledge on the architecture and implementation of VoWiFi.
In Section ~\ref{sec:methodology}, we describe our methodological approach to discover and probe the VoWiFi infrastructure of global operators.
Section~\ref{sec:results-and-evaluation} presents the results that were collected throughout this study and Section~\ref{sec:related-work} briefly outlines related studies.
Finally, we discuss our results and limitations in Section~\ref{sec:discussion} and draw final conclusions in Section~\ref{sec:conclusion}.

\section{Background}
\label{sec:background}
\begin{figure}[bt]
  \centering
  \includegraphics[width=\linewidth]{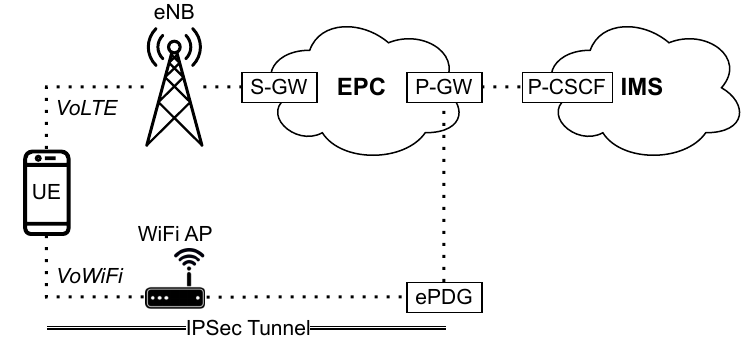}
  \caption{(Simplified) LTE network architecture for VoLTE and VoWiFi.}
  \label{fig:architecture-volte-vowifi}
\end{figure}

Figure~\ref{fig:architecture-volte-vowifi} compares VoLTE and VoWiFi within a simplified cellular network architecture.
For the sake of clarity, we've excluded several nonsubstantial components. Furthermore, we stick to the terminology that was specified in LTE, while later generations often introduced new names for similar components or services (e.g., VoLTE is called Voice over New Radio (VoNR) or Voice over 5G (Vo5G) in the fifth network generation).

\parvspace
\noindent\textbf{\textit{3GPP Access}, Voice over LTE (VoLTE).}
As shown in the upper path of Figure~\ref{fig:architecture-volte-vowifi}, the User Equipment (UE) attaches to a base station (Evolved Node B (eNB) in LTE) via the Radio Acces Network (RAN).
The Serving Gateway (S-GW) and the Packet Data Network Gateway (P-GW) within the Evolved Packet Core (EPC) system are responsible for assigning IP addresses to Access Point Names (APNs) and for routing and forwarding the traffic to external Packet Data Networks (PDNs).
Finally, the Proxy Call Session Control Function (P-CSCF) acts as a Session Initiation Protocol (SIP) proxy and is the ingress point to the IP Multimedia Subsystem (IMS). All data traffic between the UE and the P-CSCF is encapsulated in an IPsec tunnel. The UE can then directly send SIP messages, e.g., after successful establishment of the IPSec tunnel it can send a \textit{SIP REGISTER} request to the IMS core network.
While SIP is used for signaling, the actual audio stream of a call is transferred via the Real-Time Transport Protocol (RTP).

\parvspace
\noindent\textbf{\textit{Non 3GPP Access}, Voice over WiFi (VoWiFi, Wi-Fi Calling).}
In this scenario (lower path of Figure~\ref{fig:architecture-volte-vowifi}), the UE does not use the operator's RAN, but connects via an \textit{untrusted} Wi-Fi Access Point (AP). 
More specifically, it establishes another IPsec tunnel to an Evolved Packet Data Gateway (ePDG) that is accessible via the public Internet.
After successful authentication of the UE (via its IMSI and the cryptographic keys that are saved on the SIM card) and establishment of the IPSec tunnel between UE and ePDG, all traffic is forwarded to the IMS via the P-GW.
Note that for VoWiFi, the SIP traffic is actually wrapped within two different IPSec tunnels (i.e., the first between the UE and ePDG, and the second between the UE and P-CSCF).

\parvspace
\noindent\textbf{Internet Protocol Security (IPsec).}
As described above, VoLTE and VoWiFi heavily rely on IPSec~\cite{rfc6071} for authentication and traffic encapsulation.
To set up a Security Association (SA) it uses the Internet Key Exchange (IKE) protocol.
More specifically it uses IKEv2~\cite{rfc4306, rfc7296} with EAP-AKA~\cite{rfc4187} (Authentication and Key Agreement) for key derivation and thereby leverages the SIM card's secret keys to obtain a new session key.
The negotiation can be divided into two phases: \texttt{IKE\_SA\_INIT} that negotiates the ciphering suite and other security parameters, and \texttt{IKE\_AUTH} where the SIM card authenticates by solving a random challenge using its secret keys.

\section{Methodology}
\label{sec:methodology}

\begin{figure}[tb]
  \centering
  \includegraphics[width=0.765\linewidth]{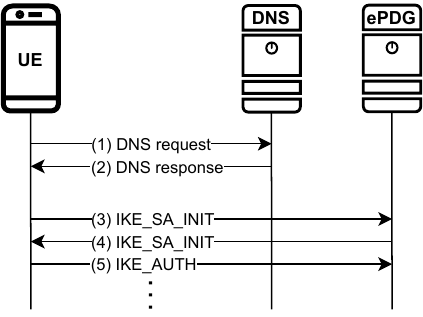}
  \caption{To connect via VoWiFi, the client fetches the ePDGs IP address via the DNS server and starts the IKE negotiation with the ePDG.}
  \label{fig:methodology-flowchart}
\end{figure}

VoWiFi calls are usually issued via domestic Wi-Fis, i.e., the customer's location can be inferred by looking at the client's IP address.
For VoWiFi, the UE needs to communicate with at least two servers, as shown in Figure~\ref{fig:methodology-flowchart}.
After discovering the responsible ePDG IPs via DNS (1, 2), the UE establishes a secure connection to the ePDG (3, 4, 5) that will be used to terminate calls and messages. %

There are multiple ways for an operator to block VoWiFi based on a subscriber's IP address:

\parvspace
\noindent\textbf{DNS.}
Global companies often use GeoDNS (also GeoIP) to minimize network latency by pointing their clients to a geographically close server.
Similarly, this can be used for geoblocking, by configuring the responsible DNS server to ignore queries that stem from unwanted client countries.
Due to caching and the recursive manner of DNS, this method is relatively inaccurate and might nevertheless leak IP addresses.
Finally, there are relatively easy anti-blocking techniques, e.g., skillful customers can use custom DNS servers or manually add host entries to circumvent geoblocking.

\parvspace
\noindent\textbf{ePDG.}
To achieve more effective blocking, operators can also implement measures at the ePDG.
As an example, an operator could simply deploy firewall rules on their ePDGs to drop packets from IPs that do not belong to the domestic country.
Additionally, operators could develop more complex rules that only permit specific \emph{premium users} (identified by their IMSI), or also check the latest roaming status (via radio) of a subscriber before deciding whether to block the connection.
To achieve worldwide coverage in our study, we decided to focus on straightforward IP-based rules because it does not require the acquisition of SIM cards of the tested operators.

To allow structured testing, our methodological approach is divided into two steps: (i) mapping DNS records and (ii) probing the actual servers.
\subsection{Mapping DNS Records}
\label{subsec:methodology:dns}
The Fully Qualified Domain Name (FQDN) of an ePDG is specified in 3GPP TS 23.003~\cite{EtsiNumberingAdressingIdentification} and can be built from an operator's Mobile Country Code (MCC) and Mobile Network Code (MNC):
\parvspace

{\hspace{0.2cm}\texttt{epdg.epc.mnc\{y\}.mcc\{x\}.pub.3gppnetwork.org}}
\parvspace
\\
According to the specification, the MCC (x) always consists of three decimal digits, while the MNC (y) can be either two or three decimal digits.
The first digit of the MCC is allocated according to the geographic region of the operator and thereby easily allows to differentiate operators based within different continents e.g., Europe or North America.

If we want to get an exhaustive list of ePDGs from all operators around the globe, requesting the IP addresses via normal (recursive) DNS requests (i.e., offloading them to popular DNS servers like Cloudflare or Google) from one central location does not work, or would at least yield imprecise or non-deterministic results (e.g., due to caching, Anycast routing, etc.).
To make recursive DNS servers query for geographically close IP addresses, the eDNS Client Subnet (ECS)~\cite{rfc7871} mechanism allows propagating the client's IP address range (usually a \texttt{/24} subnet) to the authoritative DNS server.
However, some DNS servers (e.g., Cloudflare~\cite{CloudflareFAQECS}) do not enable ECS due to privacy concerns.
Also, we found several operators' authoritative DNS servers do not support ECS and therefore solely use the request's IP address as baseline to build their responses.
Addressing this requires a more complex approach, that uses different locations worldwide to issue DNS requests in an authoritative manner.

\begin{figure}[tb]
  \centering
  \includegraphics[width=\linewidth]{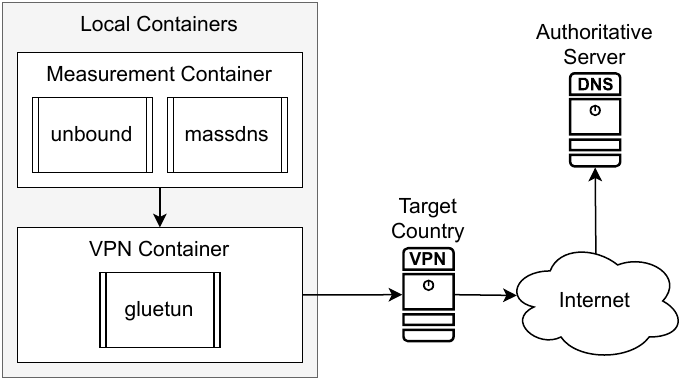}
  \caption{A containerized architecture isolates independent measurements and enables parallel execution for simple upscaling.}
  \label{fig:methodology-container-dns}
\end{figure}

To cope with these needs we use a containerized infrastructure that leverages Virtual Private Networks (VPNs) for global distribution of DNS requests.
Figure~\ref{fig:methodology-container-dns} describes the approach in detail: the VPN container connects to an existing VPN, providing Internet connectivity to the measurement container.
The measurement container runs a local \textit{unbound}\footnote{\url{https://github.com/NLnetLabs/unbound}} resolver that is configured to resolve DNS requests in an iterative manner (i.e., to get the IP addresses from the authoritative server).
All DNS requests to the local server are issued with \textit{massdns}\footnote{\url{https://github.com/blechschmidt/massdns}}.

The VPN container is built upon \textit{gluetun}\footnote{\url{https://github.com/qdm12/gluetun}}. Thereby our setup works with any \textit{OpenVPN} or \textit{WireGuard} server and already implements native support for many consumer-grade VPN services (e.g., ProtonVPN, NordVPN, CyberGhost).
To quickly achieve broad coverage, we purchase several VPN services and additionally use boto3\footnote{\url{https://github.com/boto/boto3}} to implement automatic generation of ephemeral Amazon EC2 instances that are spawned in all available AWS Regions and act as WireGuard-based relays.
Table~\ref{tab:vpn-services} provides a summary of the utilized services along with the corresponding number of countries advertised by each service.

\begin{table}[!t]%

    \newcommand\tFa{$^{\mathrm{a}}$}
    \newcommand\tNoAvail{$\times$}

    \begin{tabular}{@{}lcc@{}}
    \toprule
    Service                 & Countries\tFa         & IPv6 Support  \\ \midrule %
    Amazon EC2 (Cloud)      & 23                & \checkmark \\
    Cloudflare WARP         & 120               & \checkmark \\
    CyberGhost              & 91                & \tNoAvail \\ %
    hide.me                 & 50                & \checkmark \\
    HideMyAss               & 210               & \tNoAvail \\ %
    IVPN                    & 36                & \checkmark \\ %
    Mullvad                 & 43                & \checkmark \\ %
    NordVPN                 & 60                & \tNoAvail \\ %
    Private Internet Access & 84                & \tNoAvail \\ %
    ProtonVPN               & 68                & \tNoAvail \\ %
    Surfshark               & 100               & \tNoAvail \\ \midrule %
    \end{tabular}
    \\
    \tFa~As advertised by the VPN/cloud service.\hspace{1ex}
    
    \caption{Overview of VPN and cloud services that were used to distribute our measurements across the globe.}
    \label{tab:vpn-services}
\end{table}

\subsection{Discovering DNS Records}
\begin{table*}[t]
    \centering
    
    \newcommand\tFa{$^{\mathrm{a}}$}
    \newcommand\tFb{$^{\mathrm{b}}$}
    \newcommand\tFc{$^{\mathrm{c}}$}

\newcommand\taac{0} %
\newcommand\taad{0} %
\newcommand\taai{0} %
\newcommand\tabc{0} %
\newcommand\tabd{0} %
\newcommand\tabi{0} %
\newcommand\tacc{41} %
\newcommand\tacd{148} %
\newcommand\taci{311} %
\newcommand\tadc{20} %
\newcommand\tadd{69} %
\newcommand\tadi{127} %
\newcommand\taec{21} %
\newcommand\taed{133} %
\newcommand\taei{164} %
\newcommand\tafc{9} %
\newcommand\tafd{32} %
\newcommand\tafi{60} %
\newcommand\tagc{8} %
\newcommand\tagd{14} %
\newcommand\tagi{22} %
\newcommand\tahc{9} %
\newcommand\tahd{26} %
\newcommand\tahi{40} %
\newcommand\taic{0} %
\newcommand\taid{0} %
\newcommand\taii{0} %
\newcommand\tajc{1} %
\newcommand\tajd{1} %
\newcommand\taji{1} %
\newcommand\tatc{109}
\newcommand\tatd{423}
\newcommand\tati{725}

\newcommand\tbac{0} %
\newcommand\tbad{0} %
\newcommand\tbai{0} %
\newcommand\tbbc{0} %
\newcommand\tbbd{0} %
\newcommand\tbbi{0} %
\newcommand\tbcc{8} %
\newcommand\tbcd{9} %
\newcommand\tbci{16} %
\newcommand\tbdc{2} %
\newcommand\tbdd{2} %
\newcommand\tbdi{7} %
\newcommand\tbec{3} %
\newcommand\tbed{17} %
\newcommand\tbei{11} %
\newcommand\tbfc{1} %
\newcommand\tbfd{1} %
\newcommand\tbfi{3} %
\newcommand\tbgc{1} %
\newcommand\tbgd{1} %
\newcommand\tbgi{2} %
\newcommand\tbhc{1} %
\newcommand\tbhd{1} %
\newcommand\tbhi{1} %
\newcommand\tbic{0} %
\newcommand\tbid{0} %
\newcommand\tbii{0} %
\newcommand\tbjc{0} %
\newcommand\tbjd{0} %
\newcommand\tbji{0} %
\newcommand\tbtc{16}
\newcommand\tbtd{31}
\newcommand\tbti{40}

    \begin{tabular}{@{}lrrrrrrl@{}}
    \toprule
                                         & \multicolumn{3}{c}{IPv4}    & \multicolumn{3}{c}{IPv6} \\
    Region (via MCC)~\tFa                & Countries~\tFb & Domains & IPs   & Countries~\tFb & Domains & IPs   \\ \midrule
    0 Test networks                      & \taac     & \taad   & \taai & \tbac     & \tbad   & \tbai \\
    2 Europe                             & \tacc     & \tacd   & \taci & \tbcc     & \tbcd   & \tbci \\
    3 North America \& Caribbean         & \tadc     & \tadd   & \tadi & \tbdc     & \tbdd   & \tbdi \\
    4 Asia, Middle East                  & \taec     & \taed   & \taei & \tbec     & \tbed   & \tbei \\
    5 Australia, Oceania                 & \tafc     & \tafd   & \tafi & \tbfc     & \tbfd   & \tbfi \\
    6 Africa                             & \tagc     & \tagd   & \tagi & \tbgc     & \tbgd   & \tbgi \\
    7 South- \& Central America          & \tahc     & \tahd   & \tahi & \tbhc     & \tbhd   & \tbhi \\
    9 Worldwide~\tFc                     & \tajc     & \tajd   & \taji & \tbjc     & \tbjd   & \tbji \\ \midrule
    Total                                & \tatc     & \tatd   & \tati & \tbtc     & \tbtd   & \tbti \\ \bottomrule
    \end{tabular}
    \\
    \tFa~digits 1 and 8 are not specified.\hspace{1ex}
    \tFb~according to the MCC.\hspace{1ex}
    \tFc~Satellite, Air, Maritime, Antarctica.\hspace{1ex}\\
    \caption{Encountered ePDGs via DNS discovery, grouped by MCC region. 
    For \textit{Test networks} we found no public DNS entries.
    }
    \label{tab:found-domains-dns}
\end{table*}

To discover existing ePDGs, we simply construct a list of all possible MCC and MNC combinations, resulting in 1.1~million domain names.
We use the presented infrastructure to resolve these domain names (\texttt{A} and \texttt{AAAA} entries) from globally distributed vantage points.
For all \texttt{CNAME} responses, we iteratively resolve the referenced domain until a final response is returned.

After retrieving the corresponding \texttt{A} and \texttt{AAAA} entries for a domain we can generate an exhaustive list of all ePDGs.
For entries that only occur within specific client countries, we derive that the operator is possibly using DNS for load balancing, to reduce latency, or as a geoblocking measure.

\subsection{Probing Servers via IKE Initialization}
\label{subsec:methodology:ike}
To scan for IP-based geoblocking at the ePDG server, we can simply leverage the containerized infrastructure that was used to distribute our DNS requests in the previous section.
Within the measurement container, we run a Python script that iterates over all IP addresses that were discovered in the previous step, trying to do an \texttt{IKE\_SA\_INIT} exchange (step 3, 4 in Figure~\ref{fig:methodology-flowchart}).
The script logs its current public IP address and whether the ePDG servers respond to the sent initialization packet.
Any server that does not answer the first packet is probed repeatedly (i.e., five times) with an added backoff period.
After querying a server's status from different source IP addresses we are able to determine whether the operator drops requests that are issued from unwanted countries.

The second phase (step 5 onwards in Figure~\ref{fig:methodology-flowchart}) of the IKE protocol requires additional parameters like a subscriber's IMSI and cryptographically signed challenges that prove the identity of the customer.
Since we want to get a big picture of global geoblocking practices and because it is not feasible to get access to SIM cards of a considerable amount of all worldwide operators we focus on detecting geoblocking relying on simple rules, such as dropping packets from unwanted IP addresses.

Therefore, the results of our methodology provide a lower bound on the number of operators that deploy geoblocking for VoWiFi.
While our results in Section~\ref{sec:results-and-evaluation} show that we're able to detect a great share of blocking operators,
we outline potential factors that restrict the detection capabilities of our method in Section~\ref{subsec:limitations}.

\section{Results and Evaluation}
\label{sec:results-and-evaluation}
We started with some preliminary exploratory measurements in May 2023 and subsequently improved our measurement methodology.
The majority of our measurement results were obtained within a condensed measurement campaign during July and August 2023 (DNS discovery: Jul 13th to Aug 15th, IKE probing: Jul 13th to Aug 22th).

When citing a particular operator, we employ the notation \sloppy \texttt{CarrierName$_{[MCCMNC]}$}.

As explained in Section~\ref{subsec:methodology:dns}, we issue DNS queries for all possible domain names from multiple clients distributed worldwide.
From an abstract perspective, only a single DNS request is required to find the IP addresses that are currently associated with a domain name (from a particular location).
In practice, however, this theoretical assumption does not hold.
In fact, first of all, a client needs to find the responsible authoritative nameserver by querying the root and Top Level Domain (TLD) servers before the actual query is sent.
Additionally, when asked for an \texttt{A} or \texttt{AAAA} record, the authoritative nameserver can refer to another domain by returning a \texttt{CNAME} entry.
To cover this, we subsequently resolve \texttt{CNAME} chains until a final response (i.e., either \texttt{A}/\texttt{AAAA} or \texttt{NXDOMAIN}) is reached.
Lastly, we enable rigorous caching at our local \textit{unbound} instance, to prevent repeated queries and lower the amount of actual requests issued to external servers.

We experienced that most authoritative nameservers return a complete list of all IP addresses that are assigned to the requested domain name.
In contrast, some nameservers only answer with a single IP and withhold the rest of the addresses that are also assigned to the requested domain name.
While it is relatively easy to request all IP addresses for the first case, the latter complicates the matter and can only be tackled by querying the desired resource over and over again. %

Lastly, some nameservers are configured to return IP addresses deterministically, depending on the source of the request.%

In our approach, we split the requests into two phases (executed consecutively at every location), retrieving all available \texttt{A} and \texttt{AAAA} records respectively.

In addition to the original strategy where we query for all possible hostnames (\textit{domain discovery}), we also ran some instances only querying domains already discovered in previous rounds.
This was done to reduce the overall number of queries necessary to find all IP addresses (\textit{IP discovery}) that exist for a single domain.

Overall, we ran 
 8,555 domain discovery and 
47,902 IP discovery scans that were distributed across 219 countries.

\parvspace
\noindent\textbf{Collected IP Addresses.}
Table~\ref{tab:found-domains-dns} shows the amount of collected domains and (distinct) IP addresses.
Overall, we collected 1,026 (\texttt{A}) and 66 (\texttt{AAAA}) domain-to-IP mappings.
However, many IP addresses occur multiple times within one country, e.g., when an operator occupies multiple MNCs or when a Mobile Virtual Network Operator (MVNO) uses the ePDG of its parent provider, which reduces the set to 725 (\texttt{A}) and 40 (\texttt{AAAA}) unique ePDG IPs.
About 7.3\% of all domains support dual-stack (i.e., they have both an \texttt{A} and \texttt{AAAA} entry). For all found \texttt{AAAA} records, there is also a corresponding \texttt{A} entry, i.e., there is no operator that runs the ePDG via IPv6 only.

The vast majority of providers use three digits for the MNC in their ePDG domain, while \texttt{Vodacom$_{[64004]}$} (Tanzania) is the only one that reserves an additional two-digit domain (i.e., \sloppy
\texttt{epdg.epc.mnc04.mcc640.pub.3gppnetwork.org}), serving the same IPs as its three digit counterpart.

\parvspace
\noindent\textbf{Geographic Location of ePDGs.}
To determine the country of origin for an MNO (by its MCC), we rely on the most recent version of Android's MCC table~\cite{AOSPMCC}, which is based on T-SP-E.212A~\cite{ITUT} standardized by the International Telecommunication Union (ITU). %
To geolocate IP addresses we use the free MaxMind GeoLite2 database\footnote{\url{https://dev.maxmind.com/geoip/geolite2-free-geolocation-data}}.
The MCC country of an ePDG and the location of an ePDG according to its IP addresses do not necessarily need to be identical.
However, the majority of the operators use ePDGs that are hosted within their own network range and country.
In most cases where the ePDG is not located within the MCC country, it is hosted within close proximity (i.e., in a neighbouring country).
We noticed that this practice is especially popular with relatively small countries (e.g., \texttt{Tele2$_{[24603,24702]}$} in Latvia and Lithuania via Sweden, %
\texttt{Orange$_{[27099]}$} in Luxembourg via Belgium or \texttt{Claro$_{[74810]}$} in Uruguay via Argentina) and in Caribbean island countries (e.g., \texttt{Flow$_{[365840,364390]}$} in Anguilla and the Bahamas via Jamaica).
An outlier with no geographical proximity of MCC and ePDG country is \texttt{Tata Communications$_{[23427]}$} which is based in the United Kingdom but uses an ePDG located in the United States.
\texttt{MTX Connect$_{[90139]}$}, the only operator we found within the ``Worldwide'' MCC range, is pointing to an ePDG IP hosted in Luxembourg.
For IPv6 we do not see any divergences between MCC and ePDG country.

\parvspace
\noindent\textbf{Non-Routable IP Addresses.}
While the majority of returned IP addresses lie within public address ranges, some results are not publicly routable.
For example, German \sloppy %
\texttt{Voiceworks$_{[26220]}$} and Italian \texttt{Wind Tre$_{[22288,22299]}$}
return loopback IP addresses (\texttt{127.0.0.1} and \texttt{127.0.0.9}).
Obviously, these addresses will not work in practice and 
were presumably just deployed as a placeholder or for internal testing purposes.

For IPv6, we see several providers referring to their IPv4 siblings.
More specifically,
\texttt{Three Mobile$_{[23594]}$} (United Kingdom) and
\texttt{Méditélécom$_{[60400]}$} (Morocco)
use the NAT64 IPv6 transition mechanism~\cite{rfc6146} via the \texttt{64:ff9b::/96} prefix and
\texttt{Maxis$_{[50212]}$} (Malaysia)
refers to its IPv4 sibling via an IPv4-Compatible IPv6 address (deprecated in RFC4291~\cite{rfc4291}).

\subsection{Analyzing Differences in DNS Responses}
We experience several cases where the returned DNS results deviate for repeated queries from different locations.
For every IP address that is returned for a specific domain, we inspect the set of countries from which we were able to discover this IP address. Additionally, we also count the number of occurrences per country.

By inspecting these results, we can group the DNS servers according to their characteristic behavior:

\begin{table*}[bt]
    \begin{tabular}{@{}lrrrrl@{}}
    \toprule
                            & \multicolumn{2}{c}{IPv4}       & \multicolumn{2}{c}{IPv6}       &  \\ 
    Service                 & Countries    & Measurements    & Countries    & Measurements    &  \\ \midrule
    Amazon EC2 (Cloud)      & 21           & 2,456           & 22           & 2,212           & \\
    Cloudflare Warp         & 208          & 8,934           & 208          & 7,417           & \\
    CyberGhost              & 90           & 4,025           & 0            & 0               & \\
    hide.me                 & 49           & 1,994           & 46           & 1,641           & \\
    HideMyAss               & 207          & 2,969           & 0            & 0               & \\
    IVPN                    & 36           & 3,975           & 34           & 791             & \\
    Mullvad                 & 34           & 1,930           & 33           & 1,538           & \\
    NordVPN                 & 59           & 2,166           & 0            & 0               & \\
    Private Internet Access & 83           & 5,337           & 0            & 0               & \\
    ProtonVPN               & 68           & 3,801           & 0            & 0               & \\
    Surfshark               & 100          & 4,562           & 0            & 0               & \\ \midrule
    Total                   & 219          & 42,149          & 208          & 13,599          & \\ \bottomrule
    \end{tabular}
    \caption{To achieve global coverage and to improve diversity of our client-side vantage points, we evenly distributed the IKE probing measurements to numerous VPN- and cloud services. Surprisingly, Cloudflare Warp provided far more countries than advertised in the service description (120 advertised vs. 208 actual countries).
    While they only list locations of their data centers the service presumably also uses smaller edge locations as exit points.
    }
    \label{tab:ike-measurement-coverage}
\end{table*}

\begin{itemize}[topsep=3pt plus 2pt,itemsep=0ex, leftmargin=2em]\setlength{\itemsep}{3pt plus 2pt}\setlength{\parskip}{1pt plus 2pt minus 2pt}%

\item[\roundframe{G1}] \textit{Returning Multiple IPs for Redundancy.}~ 
This group contains all domains where the DNS server directly returns all IP addresses that are associated with the queried hostname (without differentiating by the client's location).
The vast majority of all domains (at least 78\%\footnote{According to our simple heuristic algorithm.} of all IPv4 domains) show this behavior.
Responding with a greater set of IP addresses makes sense from an availability perspective: a client that wants to connect to a service can always switch to another IP address if something goes wrong when connecting to the first endpoint.
Additionally, due to round-robin DNS~\cite{rfc1794}, this will also enforce automatic load-balancing across all associated IP addresses.

While the median number of IP addresses that are returned is only two, some operators return a greater number of IPs. For instance, the maximum number that we discovered was \texttt{Telekom$_{[26201]}$} (Germany) where all 12 IPs that exist for their ePDG domain are instantly returned by the responsible DNS server.

\item[\roundframe{G2}] \textit{Using DNS for targeted Load-balancing.}~ 
Some operators do not disclose all IP addresses in every DNS answer, but just return a subset (often just a single entry) of their responsible IP address pool.
In this case, every request receives a partition of the address pool without any specific bias or determination (regarding the request's source location).
Therefore, we assume that this is just used as a load-balancing measure, e.g., to balance clients among existing servers and that there is no intention to use this to block any resources.
Additionally, the operator could use this for A/B testing or more fine-grained balancing based on the current network status, e.g., by purposefully forwarding clients to servers that currently have free resources.

\texttt{T-Mobile$_{[310240,310260]}$} (United States) stands out as the primary example within this category, as they provide a total of 39 IP addresses for each of their domains.
All ePDG servers were in deployment concurrently (i.e., discovered during various scans in the same time period) but their DNS server only answers with a single IP per request, as described above. Supposedly, the returned IP is selected randomly, as we do not see any location bias.

\item[\roundframe{G3}] \textit{Using DNS for Geolocational Grouping.}~ 
\label{sec:dns-geological-grouping}
Again, only a single entry of a bigger address pool is returned. Additionally, the returned address is determined by the source IP address.
Analyzing the IPs returned for specific countries, we see that the operator uses predetermined IP addresses to serve customers in different locations.
For example, \texttt{Reliance Jio$_{[405874]}$}\footnote{Jio also uses 21 other MNCs that were shortened for the sake of visibility.} (India) clearly separates their domestic and foreign users.
All queries issued within India received IP addresses that never occurred within any other country. For their external customers, we could not see any further differentiation (i.e., customers in Europe usually get the same responses as customers in America or Africa).
Additionally, we noticed that their DNS server does not support the eDNS Client Subnet (ECS)~\cite{rfc7871} mechanism. Therefore, the only way to discover their local IP addresses is to issue DNS queries from a location within India.     %

Regular use cases for this behaviour include its utilization for structural grouping of customers based on their location, or reducing latency by pointing to geolocationally close servers that are close to the customers (GeoDNS). %
Additionally, this behavior could be used to block the resource by returning an IP that does not accept any connections from the requester's location.
To check whether this is the case, we need further analysis, i.e., we need to check whether the ePDG response differs between all returned IP addresses (cf. Section~\ref{sec:vowifi-geoblocking}).

\item[\roundframe{G4}] \textit{Using DNS for Geoblocking.}~ 
\label{sec:dns-geoblocking}
Finally, operators could only answer local DNS queries and simply block or drop any external requests, to prohibit their customers from accessing the IMS via Wi-Fi when being abroad.
In our results, we found that \texttt{Vodafone$_{[26202]}$} (Germany) is only serving its ePDG IP addresses for DNS requests from appropriate client IPs.
This was occasionally noticed by actual customers, as we also found anecdotal evidence~\cite{VodafoneVoWiFiIssue1,VodafoneVoWiFiIssue2,VodafoneVoWiFiIssue3,VodafoneVoWiFiIssue4,VodafoneVoWiFiIssue5,VodafoneVoWiFiIssue6}
within several posts on blogs and online forums.
In contrast to \texttt{Jio$_{[405874]}$} (see above), \texttt{Vodafone$_{[26202]}$} actually respects ECS queries, making it possible to easily demonstrate the filtering\footnote{Example queries are shown in the Appendix~\ref{sec:appendix:vodafone}}.
While most of the queries that were able to discover \texttt{Vodafone$_{[26202]}$}'s ePDG were issued within Germany, the DNS server occasionally answered requests from external countries (e.g., several close countries like Austria, and Ireland, but also more distant countries like Kazakhstan and Japan).
We tried to investigate what caused these false positives and checked the IP information of several results according to the MaxMind database.
We found, that many misclassified IPs had Germany set as the \texttt{registered\_country} (an additional meta-information in the database), although delivering an external country as the primary hosting country.
\end{itemize}

\parvspace
\noindent\textbf{IPv6.}
Again, the vast majority (over 90 \%) of all domains that return \texttt{AAAA} records can be found within the first group \roundframe{G1}.
Additionally, we found members of \roundframe{G2} and \roundframe{G3}, namely Israeli \texttt{Cellcom$_{[42502]}$}, replying with a single IP chosen without any visible location bias (\roundframe{G2}) during a transition period switching their ePDG server, and \texttt{Canadian Bell$_{[302610]}$}, that serve their customers with dedicated IPs that are (vaguely) grouped by geographic region.
In contrast to IPv4, we did not see any operator that used DNS for geoblocking purposes.

\subsection{Probing VoWiFi Servers}
\label{sec:vowifi-geoblocking}
After identifying the ePDG target IP addresses, we need to investigate whether the ePDG accepts connections from different geographic regions. 
To distribute our measurements across the globe, we again use our containerized architecture that leverages various VPN- and cloud services.

\parvspace
\noindent\textbf{Data Sources and Covered Countries.}
Both MaxMind~\cite{maxmind} and Android's MCC table~\cite{AOSPMCC} link entities (i.e., IP addresses, operators) to a geographical region (i.e., an \mbox{ISO~3166-1}~\cite{ISO3166-1} country). \mbox{ISO~3166-1} currently comprises 249 countries, and each can be linked to its corresponding continent.

In total, we issued more than 55,700 IKE scan rounds, that were executed from 219 different countries for IPv4 and 208 for IPv6 respectively.
Table~\ref{tab:ike-measurement-coverage} gives an overview on how the measurements are distributed among countries and services.

The number of countries we used for scanning greatly exceeds the number of countries where operators actually support and use VoWiFi in practice.
To maximize the scope of our study and gain a defacto worldwide view, we scanned from all 219 available countries. More specifically, this was done i) to maximize the discovered DNS entries in case of DNS-based blocking, and ii) to also detect blocking of countries that do not have VoWiFi yet, but are blocked by foreign operators (e.g., for political reasons).

\begin{figure}[t]
  \centering
  \includegraphics[width=\linewidth]{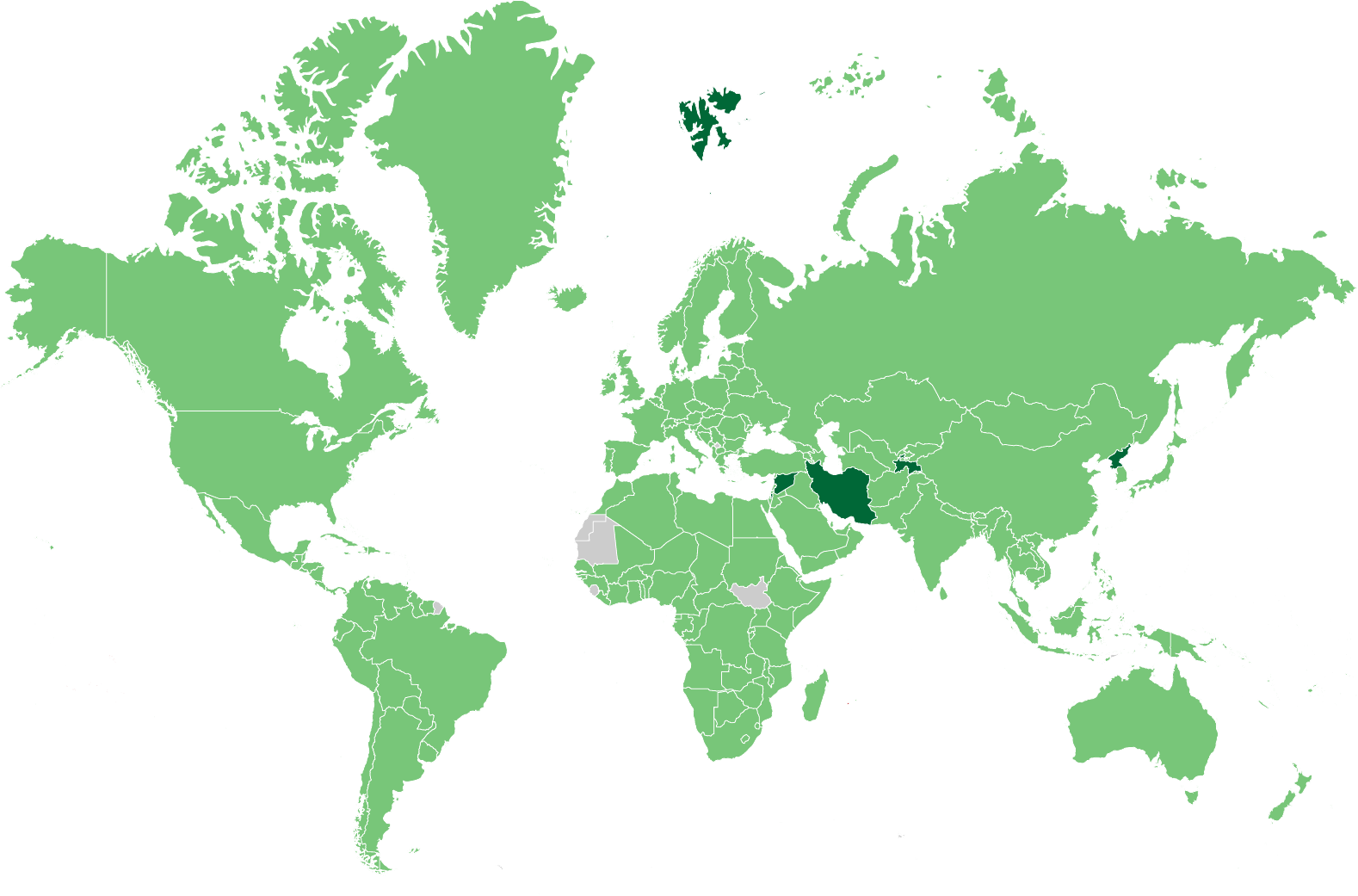}
  \textcolor[HTML]{78C679}{$\blacksquare$} Dual-stack (1)
  \hspace{1ex}
  \textcolor[HTML]{006837}{$\blacksquare$} IPv4 only (2)
  \\
  \textcolor[HTML]{CCCCCC}{$\blacksquare$} No coverage \& no VoWiFi (3)    
  \hspace{1ex}
  \textcolor[HTML]{C1272D}{$\blacksquare$} Missing coverage (4)
  \caption{Leveraging VPN and cloud services, we reach de facto worldwide measurement coverage.}
  \label{fig:image-world-coverage}
\end{figure}

\subsubsection{Measurement Coverage and Domestic Results}
Our measurements are limited by the countries that are available via our VPN- and cloud services.
We cover 107/109 (IPv4) and 16/16 (IPv6) of our target countries (target territories defined by the DNS discovery).
The two countries we are missing for IPv4 are both overseas departments of France, which are relatively small countries: French Polynesia (one operator) and La Réunion (two operators). %
Within French Polynesia, the single operator that was discovered via DNS (\texttt{Ora$_{[54705]}$}) responds to IKE requests from all over the globe and thereby is not geoblocked.
The same holds true for one Reunionese carrier(\texttt{Zeop$_{[64704]}$}), while the second one (\texttt{Orange$_{[64700]}$}) was not responsive from any country we scanned from and thereby is out of scope of our measurement coverage\footnote{For the remaining paper, we've treated it as a non-responsive ePDG, although technically it could be geoblocked and merely reachable via the home country (La Reunion).}.
Figure~\ref{fig:image-world-coverage} gives an overview of the global scope of our measurements:
We've accomplished dual-stack connectivity in nearly all covered countries~(1) and have some countries with IPv4 only coverage~(2).
Several countries that were not available via our scan infrastructure do not have VoWiFi yet~(3), i.e., there are no DNS entries for any ePDG within that area.
The two countries with missing coverage (i.e., French Polynesia and La Réunion) are very small and thereby not visible on the global map~(4).

We scan every IP address that was discovered via DNS in the previous step (Table~\ref{tab:found-domains-dns}) from all available countries.
However, not all endpoints are reachable and actually respond to requests.
Presumably, some of the DNS entries are only set for testing purposes and are in fact not in active use.
Table~\ref{tab:found-domains-ike} reduces the original DNS set by removing all IPs that do not answer requests from any possible location (i.e., not even when connecting to the ePDG from a domestic IP address).

\begin{figure}[t]
  \centering
  \includegraphics[width=\linewidth]{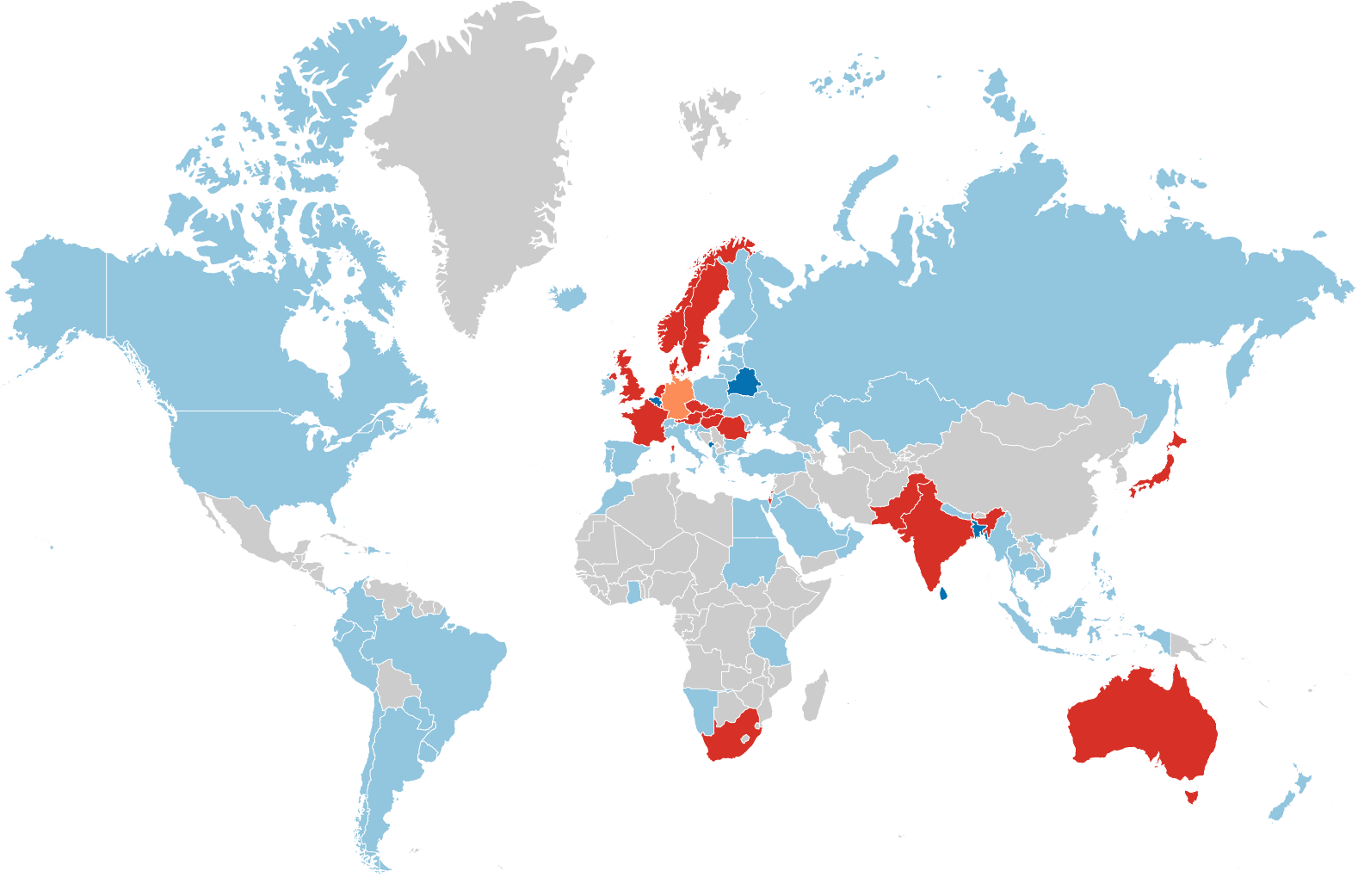}
  
  \textcolor[HTML]{92C5DE}{$\blacksquare$} No geoblocking (1)
  \hspace{1ex}
  \textcolor[HTML]{0571B0}{$\blacksquare$} DNS entry only (2)
  \\
  \textcolor[HTML]{D73027}{$\blacksquare$} Blocked[IKE] (3)     
  \hspace{1ex}
  \textcolor[HTML]{FC8D59}{$\blacksquare$} Blocked[DNS] (4)   
  \hspace{1ex}
  \textcolor[HTML]{CCCCCC}{$\blacksquare$} No VoWiFi (5)
  \caption{Summarized results or our DNS discovery and IKE probing measurements.}
  \label{fig:image-world-results}
\end{figure}

\begin{table*}[tb]
    \centering

    \newcommand\tFa{$^{\mathrm{a}}$}
    \newcommand\tFb{$^{\mathrm{b}}$}
    \newcommand\tFc{$^{\mathrm{c}}$}

\newcommand\taac{0 & -} %
\newcommand\taad{0 & -} %
\newcommand\taai{0 & -} %
\newcommand\tabc{0 & -} %
\newcommand\tabd{0 & -} %
\newcommand\tabi{0 & -} %
\newcommand\tacc{37 & 90\%} %
\newcommand\tacd{128 & 86\%} %
\newcommand\taci{271 & 87\%} %
\newcommand\tadc{19 & 95\%} %
\newcommand\tadd{61 & 88\%} %
\newcommand\tadi{113 & 89\%} %
\newcommand\taec{19 & 90\%} %
\newcommand\taed{122 & 92\%} %
\newcommand\taei{141 & 86\%} %
\newcommand\tafc{9 & 100\%} %
\newcommand\tafd{28 & 88\%} %
\newcommand\tafi{49 & 82\%} %
\newcommand\tagc{8 & 100\%} %
\newcommand\tagd{12 & 86\%} %
\newcommand\tagi{19 & 86\%} %
\newcommand\tahc{9 & 100\%} %
\newcommand\tahd{24 & 92\%} %
\newcommand\tahi{35 & 88\%} %
\newcommand\taic{0 & -} %
\newcommand\taid{0 & -} %
\newcommand\taii{0 & -} %
\newcommand\tajc{0 & 0\%} %
\newcommand\tajd{0 & 0\%} %
\newcommand\taji{0 & 0\%} %
\newcommand\tatc{101 & 93\%}
\newcommand\tatd{375 & 89\%}
\newcommand\tati{628 & 87\%}

\newcommand\tbac{0 & -} %
\newcommand\tbad{0 & -} %
\newcommand\tbai{0 & -} %
\newcommand\tbbc{0 & -} %
\newcommand\tbbd{0 & -} %
\newcommand\tbbi{0 & -} %
\newcommand\tbcc{5 & 62\%} %
\newcommand\tbcd{5 & 56\%} %
\newcommand\tbci{9 & 56\%} %
\newcommand\tbdc{1 & 50\%} %
\newcommand\tbdd{1 & 50\%} %
\newcommand\tbdi{5 & 71\%} %
\newcommand\tbec{3 & 100\%} %
\newcommand\tbed{16 & 94\%} %
\newcommand\tbei{8 & 73\%} %
\newcommand\tbfc{0 & 0\%} %
\newcommand\tbfd{0 & 0\%} %
\newcommand\tbfi{0 & 0\%} %
\newcommand\tbgc{0 & 0\%} %
\newcommand\tbgd{0 & 0\%} %
\newcommand\tbgi{0 & 0\%} %
\newcommand\tbhc{1 & 100\%} %
\newcommand\tbhd{1 & 100\%} %
\newcommand\tbhi{1 & 100\%} %
\newcommand\tbic{0 & -} %
\newcommand\tbid{0 & -} %
\newcommand\tbii{0 & -} %
\newcommand\tbjc{0 & -} %
\newcommand\tbjd{0 & -} %
\newcommand\tbji{0 & -} %
\newcommand\tbtc{10 & 62\%}
\newcommand\tbtd{23 & 74\%}
\newcommand\tbti{23 & 57\%}

    \begin{tabular}{@{}lrr|rr|rr||rr|rr|rrl@{}}
    \toprule
                                         & \multicolumn{6}{c}{IPv4}    & \multicolumn{6}{c}{IPv6} \\
    Region (via MCC)                & \multicolumn{2}{c}{Countries~\tFb} & \multicolumn{2}{c}{Domains} & \multicolumn{2}{c}{IPs}  & \multicolumn{2}{c}{Countries~\tFb} & \multicolumn{2}{c}{Domains} & \multicolumn{2}{c}{IPs}   \\ \midrule
    2 Europe                             & \tacc     & \tacd   & \taci & \tbcc     & \tbcd   & \tbci \\
    3 North America \& Caribbean         & \tadc     & \tadd   & \tadi & \tbdc     & \tbdd   & \tbdi \\
    4 Asia, Middle East                  & \taec     & \taed   & \taei & \tbec     & \tbed   & \tbei \\
    5 Australia, Oceania                 & \tafc     & \tafd   & \tafi & \tbfc     & \tbfd   & \tbfi \\
    6 Africa                             & \tagc     & \tagd   & \tagi & \tbgc     & \tbgd   & \tbgi \\
    7 South- \& Central America          & \tahc     & \tahd   & \tahi & \tbhc     & \tbhd   & \tbhi \\
    9 Worldwide~\tFc                     & \tajc     & \tajd   & \taji & \tbjc     & \tbjd   & \tbji \\ \midrule
    Total                                & \tatc     & \tatd   & \tati & \tbtc     & \tbtd   & \tbti \\ \bottomrule
    \end{tabular}
    \\
    \tFb~according to the MCC.\hspace{1ex}
    \tFc~Satellite, Air, Maritime, Antarctica.\hspace{1ex}\\
    \caption{Overview of ePDGs responding to our IKE scan from at least one vantage point. The percentage columns compare the values to the results of the previous step (i.e., the values in Table~\ref{tab:found-domains-dns}).}
    \label{tab:found-domains-ike}
\end{table*}

\parvspace
\noindent\textbf{Dealing with Inaccuracies.}
Besides blocking according to a user's location, operators might also block the usage of popular VPN services or cloud infrastructure as a connection relay.
More specifically, they might block some of our VPN services, or Autonomous Systems (ASes) that are commonly shared among popular VPN services.
Furthermore, not all operators use MaxMind as their geolocation broker, introducing potential inconsistencies between our classification and the actual geolocation at the operator's sides.
Additionally, we might occasionally experience connection hiccups within our scan infrastructure (e.g., when the used VPN unexpectedly closes the current connection).
Lastly, some ePDGs seem to be very sensitive and do not accept repeated connections from the same IP address, simply ignore IKE handshakes that contain inappropriate cipher sets or go on/offline (i.e., experience downtimes) throughout our measurement study.
To cope with these instabilities, we randomly switch VPN servers between subsequent measurements and aim for a big and diverse set of VPN services and realized measurements.

Also, we use the \textit{responsiveness} (i.e., the percentage an ePDG was reported as responsive) of the domestic case as a \textit{baseline}, to decide whether the server was blocked in a roaming case.
As a matter of precaution, we consider some space for generous error bounds (e.g., when we or the operator misclassify the VPN IP and assume the wrong country) by only flagging an area as \textit{geoblocked}, when it has less than 10\% of the assumed \mbox{\textit{baseline responsiveness}}.
Thereby, the responsiveness remains comparable and we can correctly classify an ePDG, even when unexepcted events occur, e.g., when the ePDG server experiences a downtime during our measurement campaign.
Also, this approach allows us to more accurately flag a country as \textit{geoblocked}, even when we get responses in a minority of cases due to geolocation errors from the operator.

We do not differentiate between a successful \texttt{IKE\_INIT} phase and a received error message.
For example, when the ePDG responds with a \texttt{NO\_PROPOSAL\_CHOSEN} error to indicate that the offered
ciphers are not accepted by the gateway we assume that there is no geoblocking for this client IP since a response packet was received (i.e., there is no IP-based blocking for this location).

\subsubsection{Roaming Results}
We found geoblocking at the IKE layer in various forms and granularities.
For example, the target area of the blocking can be rather big (e.g., blocking all foreign connections) or small (e.g., only blocking a specific set of target countries).
The confidence and robustness of our classification results (whether a server blocks by location or not) corresponds to the amount and diversity (e.g., IP- and AS diversity) of our measurements within that area.
Therefore, we automatically classify global and continental geoblocking, but rely on manual inspection to identify country-level geoblocking, due to reduced diversity in small target countries.
A domain is only flagged as geoblocked, when we observe geoblocking for all corresponding IP addresses.

\parvspace
\noindent\textbf{Large-scale Geoblocking.}
According to our results, we experience global geoblocking for 12.5\% (IPv4) and 65.2\% (IPv6) of all tested domains.
If we extend this to domains that are blocked from at least one continent (while being reachable from other areas) the percentage increases to  14.6\% for IPv4 and remains unchanged for IPv6.
Table~\ref{tab:ike-blocked} shows the detailed roaming results of our IKE probing.
Overall, our measurements show that geoblocking is especially common within Europe and Asia but there are also continents without any large-scale blocking at all (e.g., North and South America).

Interestingly, we found numerous operators in European countries (Austria, Czech Republic, Denmark, France, Hungary, Luxembourg, the Netherlands, Norway, Romania, Sweden, and the United Kingdom) with large-scale blocking measures.
Moreover, many operators within the EU also block connections from their neighboring EU countries.
In contrast, intra-EU roaming via regular radio access is possible without additional costs in these countries, due to the Roam Like At Home (RLAH) doctrine~\cite{EU-2022-612}.
As an exception, we've found a single Slovakian operator with worldwide geoblocking that specifically exempts EU/EAA countries from the blocking. %

While most operators employing large-scale blocking within Asia are based in India (15/37), we've also discovered geoblocking at operators from Hong Kong, Israel, Japan, Pakistan, and Singapore.
Lastly, the remaining operators were found in Australia and South Africa.

The overall results are visualized in Figure~\ref{fig:image-world-results}.
For the majority of countries there was no large-scale geoblocking discovered (1).
Some territories had DNS entries for an ePDG, but did not respond to any of our IKE scans (2).
Moreover, we found geoblocking measures via IKE (3) and DNS (4) scans.
Finally, some countries do not have DNS entries for ePDGs at all and therefore do not support VoWiFi yet~(5).

\begin{table*}[tb]
    \centering

    \newcommand\tFa{$^{\mathrm{a}}$}
    \newcommand\tFb{$^{\mathrm{b}}$}
    \newcommand\tFc{$^{\mathrm{c}}$}

\newcommand\taac{0 & -} %
\newcommand\taad{0 & -} %
\newcommand\taai{0 & -} %
\newcommand\tabc{0 & -} %
\newcommand\tabd{0 & -} %
\newcommand\tabi{0 & -} %
\newcommand\tacc{12 & 32\%} %
\newcommand\tacd{19 & 15\%} %
\newcommand\taci{35 & 13\%} %
\newcommand\tadc{0 & 0\%} %
\newcommand\tadd{0 & 0\%} %
\newcommand\tadi{0 & 0\%} %
\newcommand\taec{5 & 26\%} %
\newcommand\taed{32 & 26\%} %
\newcommand\taei{33 & 23\%} %
\newcommand\tafc{2 & 22\%} %
\newcommand\tafd{3 & 11\%} %
\newcommand\tafi{6 & 12\%} %
\newcommand\tagc{1 & 12\%} %
\newcommand\tagd{1 & 8\%} %
\newcommand\tagi{2 & 11\%} %
\newcommand\tahc{0 & 0\%} %
\newcommand\tahd{0 & 0\%} %
\newcommand\tahi{0 & 0\%} %
\newcommand\taic{0 & -} %
\newcommand\taid{0 & -} %
\newcommand\taii{0 & -} %
\newcommand\tajc{0 & -} %
\newcommand\tajd{0 & -} %
\newcommand\taji{0 & -} %
\newcommand\tatc{20 & 20\%}
\newcommand\tatd{55 & 15\%}
\newcommand\tati{76 & 12\%}

\newcommand\tbac{0 & -} %
\newcommand\tbad{0 & -} %
\newcommand\tbai{0 & -} %
\newcommand\tbbc{0 & -} %
\newcommand\tbbd{0 & -} %
\newcommand\tbbi{0 & -} %
\newcommand\tbcc{0 & 0\%} %
\newcommand\tbcd{0 & 0\%} %
\newcommand\tbci{0 & 0\%} %
\newcommand\tbdc{0 & 0\%} %
\newcommand\tbdd{0 & 0\%} %
\newcommand\tbdi{0 & 0\%} %
\newcommand\tbec{2 & 67\%} %
\newcommand\tbed{15 & 94\%} %
\newcommand\tbei{7 & 88\%} %
\newcommand\tbfc{0 & -} %
\newcommand\tbfd{0 & -} %
\newcommand\tbfi{0 & -} %
\newcommand\tbgc{0 & -} %
\newcommand\tbgd{0 & -} %
\newcommand\tbgi{0 & -} %
\newcommand\tbhc{0 & 0\%} %
\newcommand\tbhd{0 & 0\%} %
\newcommand\tbhi{0 & 0\%} %
\newcommand\tbic{0 & -} %
\newcommand\tbid{0 & -} %
\newcommand\tbii{0 & -} %
\newcommand\tbjc{0 & -} %
\newcommand\tbjd{0 & -} %
\newcommand\tbji{0 & -} %
\newcommand\tbtc{2 & 20\%}
\newcommand\tbtd{15 & 65\%}
\newcommand\tbti{7 & 30\%}

    \begin{tabular}{@{}lrr|rr|rr|rrl@{}}
    \toprule
                                         & \multicolumn{4}{c}{IPv4}    & \multicolumn{4}{c}{IPv6} \\
    Region (via MCC)                     & \multicolumn{2}{c}{Countries~\tFb} & \multicolumn{2}{c}{Domains} & \multicolumn{2}{c}{Countries~\tFb} & \multicolumn{2}{c}{Domains}   \\ \midrule
    2 Europe                             & \tacc     & \tacd   & \tbcc     & \tbcd   \\
    3 North America \& Caribbean         & \tadc     & \tadd   & \tbdc     & \tbdd   \\
    4 Asia, Middle East                  & \taec     & \taed   & \tbec     & \tbed   \\
    5 Australia, Oceania                 & \tafc     & \tafd   & \tbfc     & \tbfd   \\
    6 Africa                             & \tagc     & \tagd   & \tbgc     & \tbgd   \\
    7 South- \& Central America          & \tahc     & \tahd   & \tbhc     & \tbhd   \\ \midrule
    Total                                & \tatc     & \tatd   & \tbtc     & \tbtd   \\ \bottomrule
    \end{tabular}
    \\
    \tFb~according to the MCC.\hspace{1ex}
    \caption{Overview of ePDGs where we encountered large-scale (global or continental) geoblocking. The percentage columns relate the values to the parent population of responsive ePDGs in the corresponding area (cf. Table~\ref{tab:found-domains-ike}).}
    \label{tab:ike-blocked}
\end{table*}

\parvspace
\noindent\textbf{Country-targeted Geoblocking.}
In some cases we see more fine-grained blocking or exemptions from large-scale blocks.
For example, one Australian operator that generally blocks foreign IPs specifically allows connections from other
Oceanian (e.g., New Zealand) and
Asian (e.g., the Philippines, Malaysia , China)
countries.

However, we also see country-targeted blocking within continents and countries that otherwise do not engage geoblocking.
For example, an Israeli operator specifically blocks several
African (e.g., Lybia, Algeria) and
Asian (e.g., Iran, Iraq)
countries, despite their geographical proximity.
Moreover, several operators, e.g., in the United States or Ecuador, block connections coming from Russia or Ukraine, potentially for political or security reasons.

\parvspace
\noindent\textbf{IPv6.}
For IPv6, we've detected geoblocking for 65.2\% of all tested domains.
However, these numbers are only caused by operators from two distinct countries: India and Japan.
Compared to IPv4, the overall percentage is higher because India is among the leading nations when it comes to IPv6 adoption and simultaneously a country where geoblocking at VoWiFi is fairly popular.

Interestingly, we've found one Hungarian operator that supports both IPv4 and IPv6 and blocks external connections only on the IPv4 stack.
Thereby, customers could circumvent the blocking by simply connecting via IPv6 (a similar phenomenon has been discovered by previous research ~\cite{czyz2016backdoor}).

Overall, the available geolocation data seems to be more accurate for IPv6 because --- compared to IPv4 --- we see less blurring (i.e., for each distinct country the responsiveness is either 0 or 100\%).

\parvspace
\noindent\textbf{Revisited: \texttt{Reliance Jio$_{[405874]}$} India.}
As stated in Section~\ref{sec:dns-geological-grouping}, we found that this operator clearly separates the IP addresses that are returned via DNS for local and external customers.
Analyzing the two subsets (i.e., local, and external) at the ePDG layer, we see that this behavior also reoccurs when connecting to the gateway:
The set of local IP addresses does not accept any connections from abroad. Interestingly, this behavior is also mutual, i.e., the external ePDG does not accept any local connections from India either.
Since a customer can thereby establish a connection to an ePDG from any location we did not account this behaviour as geoblocking.
However, the existing separation mechanism could possibly be used to differentiate or block at a later stage.

\parvspace
\noindent\textbf{Revisited: \texttt{Vodafone$_{[26202]}$} Germany.}
In Section~\ref{sec:dns-geoblocking} we showed that this operator uses DNS-based blocking to prevent customers from connecting to the ePDG from external locations.
Taking a closer look at the probing results of the corresponding endpoints, we see that the blocking is solely done by the DNS server and does not occur at the IKE layer (i.e., it accepts connections from all tested countries).
Thereby, sneaky customers may simply evade the blocking by manually adding the correct DNS entries to their system, or by using specific DNS servers that always return the corresponding IPs (e.g., a local resolver in Germany that does not forward the client's subnet).

\section{Related Work}
\label{sec:related-work}
This section presents an overview of existing research and studies that contribute to the understanding and context of the subject matter at hand.

\parvspace
\noindent\textbf{Geoblocking and Internet Censorship.}
In 2018, the European Union (EU) banned unjustified geoblocking within the European single market~\cite{EU-2018-302}.
Nevertheless, the regulation includes a number of exceptions (e.g., for copyrighted audiovisual content like Netflix) and significant portions of the world remain without regulatory measures.

McDonald et al.~\cite{McDonald2018Forbidden} proved the prevalence of geoblocking practices in the Internet by finding geoblocking at large CDNs for nearly all of the 177 examined countries.

Additionally, Kumar et al.~\cite{Kumar2022Large} analyzed the mobile app ecosystem from vantage points in 26 countries.
Aside from geoblocking being a common practice in the mobile app field they also found geodifferences occurring between differing countries (i.e., developers shipping different versions of an app to specific regions).

Lastly, geoblocking is often introduced as a governmental censorship measure.
Ramesh et al.~\cite{Ramesh2023Network} showed, that --- ever since Russia's invasion of Ukraine in February 2022 --- there are geofences between Russia and the rest of the world. Thereby, Russian users are not able to consume Western news or social media and Russian government domains remain inaccessible from regions like the EU and US.

\parvspace
\noindent\textbf{VoLTE and VoWiFi Security.}
Prior research discovered numerous security- and privacy-related vulnerabilities in VoLTE and VoWiFi.
For example, Kim~\cite{Kim2015Breaking} showed that early VoLTE deployments were prone to data traffic free-riding attacks since the packet-switched voice channel provided an unmetered breakout to the public Internet.
Furthermore, both VoLTE and VoWiFi were found to occasionally leak precise subscriber info (e.g., Cell IDs) via the underlying SIP traffic~\cite{Kim2015Tracking}.
Additionally, VoWiFi is vulnerable to IMSI catching attacks~\cite{OHanlon2016Blackhat, OHanlon2017Mobile}.

More recently, Lu et al.~\cite{Lu2020Ghost}, Xie et al.~\cite{Xie2020Untold}, and Lee et al.~\cite{Lee2022VWAnalyzer} presented practical Denial of Service (DoS) attacks for VoLTE and/or VoWiFi. Moreover, Hu et al. showed that VoLTE's emergency services are also vulnerable to DoS and free-riding attacks~\cite{Hu2022Uncovering}.

In 2023, the Google Project Zero team discovered four severe Exynos vulnerabilities that allowed an attacker to execute arbitrary commands on the baseband processor of the most recent Pixel and many Samsung phones by injecting malicious SIP messages~\cite{Google2023Exynos} into the VoLTE/VoWiFi VoIP traffic.

Lastly, Gegenhuber et al.~\cite{Gegenhuber2024Never, Gegenhuber2024Diffie} uncovered insecure VoWiFi configurations and shortcomings at the corresponding key exchange.

\parvspace
\noindent\textbf{Active Roaming Experiments.}
Large-scale studies that involve many operators and countries are usually limited by the complex ecosystem and the required coordination effort. However, there are several approaches~\cite{Mandalari2018Experience, Varvello2023Worldwide} where specifically built measurement devices were placed into target locations to measure the implications (e.g., QoE) of roaming.
Additionally, Sahin and Francillon~\cite{Sahin2016Over} observed hijacking of traditional voice calls that were redirected to over-the-top (OTT) services (e.g., WhatsApp, Viber) to bypass/monetize termination fees.

Recently, Gegenhuber et al.~\cite{Gegenhuber2023MobileAtlas, Gegenhuber2022Zero} introduced the MobileAtlas measurement platform that tries to overcome the mentioned scalability issues by tunneling the communication between SIM card and modem over the Internet. Their platform provides flexible roaming measurements and capabilities for a rich set of cellular features, including Internet and voice-based measurements.

\parvspace
\noindent\textbf{Evaluating and Fingerprinting VPNs.}
The commercial VPN ecosystem is a multi-billion dollar industry~\cite{VPNDemand} and thereby has been an interesting research target.
For example, previous work~\cite{Khan2018Empirical, Ramesh2022Vpnalyzer} analyzed and compared existing VPN services and found that many solutions leak user traffic or advertise wrong server locations.

In contrast, Maghsoudlou et al.~\cite{Maghsoudlou2023Characterizing} did not purchase any VPN subscriptions but executed Internet-wide scans to discover and fingerprint VPNs, finding over 7 million IPsec servers.

\section{Discussion}
\label{sec:discussion}

Our findings indicate that a notable proportion of operators are implementing IP-based restrictions to prevent customers from using VoWiFi in specific locations.
While we also discovered DNS-based approaches, most operators implement it directly at the IKE layer.
Furthermore, the found geoblocking measures exhibit a degree of regionality, meaning they are more prevalent in certain areas (e.g., Eurasia) compared to others (e.g., North- and South America).

To the best of our knowledge, this is the first study providing a comprehensive overview of the existing VoWiFi infrastructure on a global scale.
Understanding the current deployment of any widely used telecommunication system is vital from a security standpoint. 
Our findings go beyond this by revealing that the observed blocking has significant repercussions for emergency calling.

\parvspace
\noindent\textbf{Implications to Emergency Calling.}
Recent reports show, that there currently is no adequate support for VoLTE roaming in substantial parts of the world~\cite{Capacitymedia2022VoLTE1, Capacitymedia2022VoLTE2}.
Since many operators are actively shutting down their 2G/3G legacy networks, this scenario has the potential to result in significant repercussions for the functionality of emergency calling~\cite{Hendrik2022VoLTE, EENA2022VoLTE, EENA2022Webinar}.

Although we believe that affected operators will address and fix these issues in the long term, VoWiFi could help to mitigate these shortcomings in the present day.
Tourists or travellers frequently have access to WiFi, such as in their accommodations or through free WiFi hotspots in public spaces.
According to the specification~\cite{EtsiArchitectureNon3gpp}, VoWiFi should be used by the UE for emergency calling when traditional radio services (VoLTE roaming, CSFB) are unavailable.
The phone ultimately tries to reach both the home and the visited ePDG. However, there are circumstances where the visited ePDG might not be available, e.g., when there is no VoWiFi support in the visited country or when the customer leaves the phone in flight mode to not cause any unintentional roaming fees and thereby the current location is not known to the phone).

While operators can override the default ePDG for emergency services by setting corresponding DNS records (\texttt{sos.epdg.epc.mnc\{y\}.mcc\{x\}.pub.3gppnetwork.org})~\cite{EtsiNumberingAdressingIdentification}, only four operators had appropriate DNS entries. In all four cases, the emergency ePDG referenced the original ePDG's IP address, which is also the default behaviour when no \texttt{sos} entry is found.
Controversially, two of the operators specifically using \texttt{sos}-domains nevertheless block IKE-inquiries coming from foreign IP addresses.

If an operator deploys DNS- or IKE-based geoblocking, these measures will also impact emergency services, actively denying persons in need from making an emergency call.
Similarly, the emergency service via Wi-Fi is also unavailable in the home country, when customers use a VPN connection or an international SIM card (e.g., utilizing a travel router) that provides the Internet uplink via a foreign country.

\parvspace
\noindent\textbf{Economic and Net Neutrality Perspective.}
In contrast to regular roaming over the radio interface -- where the foreign roaming partner charges the home operator for the terminated calls and services -- there is no additional economic overhead for VoWiFi calls that are initiated from overseas customers.
In fact, calls terminated via VoWiFi are notably cost-effective for operators, as they eventually reduce expenses associated with the required radio transmission infrastructure (i.e., base stations) and spectrum licensing fees.
Instead, the traffic is routed via external infrastructure (i.e., a WiFi AP) that was provided and paid for by the customer.

Moreover, adhering to the principle of net neutrality and the Open Internet, it is not allowed to discriminate (i.e., block) a customer's data packets by their source or destination IP address, particularly when motivated solely by economic interests (cf. differential pricing and zero-rating).
Assuming the discovered blocking practices were employed mainly for economic reasons, they could potentially be seen as a net neutrality violation.

\parvspace
\noindent\textbf{Ways to Evade Geoblocking.}
Common mobile operating systems (i.e., Android and iOS) support changing the used DNS server for WiFi interfaces out of the box, which should allow easy bypassing of DNS-based blocking.

Additionally, both Android and iOS have built-in support for applying a system-wide VPN connection.
However, even with an active VPN, VoWiFi uses a direct route over the WiFi interface.
While this hinders standalone solutions on non-rooted phones, a viable alternative could be to use so-called travel routers. These devices act as an intermediary between WiFi AP and smartphone and typically offer the ability to redirect all traffic through a VPN connection, and thus via a customer's home country.

\parvspace
\noindent\textbf{Inaccurate Geolocation and Blocked VPN Services.}
Each VPN service uses one or more IP addresses for each country within its coverage.
In the course of this study, we occasionally experienced outliers where certain IP addresses, address ranges, or Autonomous Systems (ASes) from specific VPNs experienced blocking even though belonging to the same country as the probed operator.
We suppose that those connections were either blocked because their geolocation was wrongfully classified from the operator, or because the operator specifically blocked connections from IPs that are associated with certain VPN services.

\subsection{Limitations}
\label{subsec:limitations}
Although we are able to detect a considerable amount of blocking operators, several factors limit our approach.
Due to the fact that these constraints limit the geoblocking variants that we're able to detect, our results can be seen as a lower bound.
Thereby, the actual occurrence of geoblocking in practice is most likely even higher.

\parvspace
\noindent\textbf{More Advanced Blocking.}
Our current approach is designed to detect simple IP-based blocking rules.
However, some operators use more sophisticated (e.g., IMEI-based) ways to decide whether a customer should be able to use VoWiFi under current circumstances.
For example, an operator could use the last known roaming status via the radio network to decide whether a customer is currently in their home country or abroad.
Additionally, VoWiFi roaming can only be offered to a limited set of customers (i.e., \textit{premium users}).
For example, we found an Austrian operator that requires an additional subscription package to get VoWiFi enabled during international roaming~\cite{Magenta2023VoWiFiRoaming}.
In this case, the operator initially accepts IKE packets from all locations and decides whether to grant (or drop) access to VoWiFi at a later stage, when the subscriber's identity is known.
We focus on straightforward IP-based blocking because measuring these more advanced blocking techniques on a global scale would necessitate the acquisition of SIM cards for an unfeasible number of operators.
Moreover, it is crucial to highlight that relying solely on IP address-based blocking for VoWiFi access also results in the restriction of access to emergency calling services.
Implementing blocking techniques that compromise the unconditional availability of emergency services that people depend on in life-threatening situations is highly ill-advised.

\parvspace
\noindent\textbf{Implementation Incompatibility and Blocked VPN Services.}
Our IKE probing is based on a cellular-specific open-source implementation of the IKEv2 protocol\footnote{\url{https://github.com/fasferraz/SWu-IKEv2}}.
Possibly, some ePDGs are not compatible with this implementation or do not answer requests that offer inappropriate cipher suites.
Similarly, operators could block access from well-known VPN services or ASes and IP ranges that are used by them.
As a countermeasure, we use multiple services to increase the diversity within our clients.
Since we got responses from a high share of ePDGs that were discovered via DNS, these two factors do not considerably influence our results.

\parvspace
\noindent\textbf{Non-Native and Non-3GPP-Compliant VoWiFi.}
Whereas our approach focuses on the native 3GPP version of VoWiFi there are also other ways to support voice calling functionalities over Wi-Fi networks.
For example, an operator might require their customers to download and install a dedicated VoWiFi app that gives direct access to the IMS network in a non-compliant way~\cite{GSMA2017ChinaVoWiFi}.
Additionally, some operators might use non-default ways to communicate their ePDGs, e.g., only resolve the hostname via an internal DNS server that is shipped to their customers via DHCP.
Lastly, we do not cover VoIP calling via OTT services (e.g., WhatsApp, Viber, Skype).

\subsection{Ethical Considerations}
Ethical considerations are vital to the field of measurements, especially with active measurements conducted in live systems.
From an operator's perspective, we only interact with two endpoints i.e., retrieving IP addresses via the authoritative DNS server and performing the initial \texttt{IKE} handshake with the ePDG.
In both cases, we tried to mimic normal user behavior by sending well-formed requests, i.e., we did not deviate from the protocol specification or do any fuzzing.
From a bandwidth perspective, our measurements should not overstress any operator, since for each measurement target we were only sending several requests per hour.
Finally, our measurements did not involve any actual user data (e.g., IMSIs or IMEIs), since we were only performing the initial handshake where the cryptographic parameters for the subsequent connection are exchanged.

\subsection{Dissemination and Responsible Disclosure}
Beyond disseminating our findings to the scientific community, we want to enable an informed debate by raising awareness within the industry and
highlight potential issues regarding IP-based geoblocking among regulators and emergency associations.
Therefore, we reported our results to the GSM Association (GSMA), the Body of European Regulators for Electronic Communications (BEREC), and the European Emergency Number Association (EENA).
GSMA and EENA invited us to expand upon and discuss our findings within dedicated meetings and BEREC responded to our inquiry via Email.

More specifically, GSMA discussed the topic within their panel of experts and invited us to \textit{``present this case to the GSMA's Fraud and Security Architecture Group, to make the issue know to their members''}.
According to BEREC \textit{``there are no legal obligations for Wi-Fi calls in roaming stemming from the Roaming Regulation, and they do not see this as a breach of the Open Internet Regulation''}.
Lastly, \textit{``EENA will study the topic and discuss with its community to understand the extent of any potential impact the practice of geoblocking could have on access to emergency services through emergency communications''}.

\section{Conclusion}
\label{sec:conclusion}
Many operators worldwide have implemented support for VoLTE and VoWiFi and rely on the IP Multimedia Subsystem (IMS) as a centerpiece for their communication services.
Additionally, the IMS and VoWiFi seamlessly integrate with the upcoming 5G network generation and thereby will also play a relevant role in the future.

Just like any crucial system that impacts our daily lives, it's important to be aware of its current state and to comprehend its internal workings.
We therefore give a comprehensive overview of the global deployment of VoWiFi and investigate existing geoblocking measures, discovering IP-based blocking mechanisms both at the DNS and IKE layer.
We emphasize that, unlike geoblocking measures commonly employed in web or streaming applications, telecommunications is a more sensitive domain, where such measures could potentially have adverse effects on the functionality of emergency calls.
Thus, we hope that the insights of our study will raise the awareness among customers and security researchers, while also contributing to the decision-making processes of policy makers and operators in the future.

To encourage other researchers to further profit from our work and engineering effort, we've publicly released the source code of our measurement infrastructure~\footnote{\url{https://github.com/sbaresearch/scanywhere}}.

\section*{Acknowledgements}
We want to thank Quentin McGaw and all other Gluetun contributors
-- their previous work made it considerably easier to build \textit{scanywhere} and thus to globally distribute our measurements.

This project was funded through the NGI0 Entrust Fund, a fund established by NLnet with financial support from the European Commission's Next Generation Internet, under the aegis of DG Communications Networks, Content and Technology under grant agreement No 101069594.

SBA Research (SBA-K1) is a COMET Centre within the COMET -- Competence Centers for Excellent Technologies Programme and funded by BMK, BMAW, and the federal state of Vienna. COMET is managed by FFG.

\appendix
\section{Appendix}
\label{sec:appendix}

\subsection{DNS-based Blocking at Vodafone}
\label{sec:appendix:vodafone}
\noindent Resolving standardized ePDG domain to \texttt{CNAME} reference:
\begin{footnotesize}
\begin{verbatim}
$ dig epdg.epc.mnc002.mcc262.pub.3gppnetwork.org
=> returns CNAME epdg.epc.drz1.vodafone-ip.de
\end{verbatim}
\end{footnotesize}

\noindent Actual resolution (Google vs. Vodafone IP range):
\begin{footnotesize}
\begin{verbatim}
# requesting via Google IP (United States)
$ dig +trace epdg.epc.drz1.vodafone-ip.de +subnet=104.154.0.0/24
\end{verbatim}
\end{footnotesize}

\begin{footnotesize}
\begin{verbatim}
# requesting via Vodafone IP (Germany)
$ dig +trace epdg.epc.drz1.vodafone-ip.de +subnet=109.192.0.0/24
\end{verbatim}
\end{footnotesize}

\subsection{Artifact Appendix}
The research artifacts accompanying this paper are available via \href{https://doi.org/10.5281/zenodo.11089362}{10.5281/zenodo.11089362}.

\printbibliography

\end{document}